\documentclass[twocolumn,showpacs,preprintnumbers,amsmath,amssymb,prb]{revtex4}

\usepackage{graphicx}
\usepackage{dcolumn}
\usepackage{bm}
\usepackage{color}
\begin{document}

\bibliographystyle{naturemag}

\preprint{}

\title{Engineering near infrared single photon emitters in ultrapure silicon carbide}

\author{F.~Fuchs$^{1}$}
\email[These authors contributed equally to this work.]{}
\author{B.~Stender$^{1}$}
\email[These authors contributed equally to this work.]{}
\author{M.~Trupke$^{2}$}
\author{J.~Pflaum$^{1,3}$}
\author{V.~Dyakonov$^{1,3}$}
\email[E-mail:~]{dyakonov@physik.uni-wuerzburg.de}
\author{G.~V.~Astakhov$^{1}$}
\email[E-mail:~]{astakhov@physik.uni-wuerzburg.de}

\affiliation{$^1$Experimental Physics VI, Julius-Maximilian University of W\"{u}rzburg, 97074 W\"{u}rzburg, Germany  \\
$^2$Vienna Center for Quantum Science and Technology, Atominstitut, TU Wien, 1020 Wien, Austria\\
$^3$Bavarian Center for Applied Energy Research (ZAE Bayern), 97074 W\"{u}rzburg, Germany}


\date{\today}


\maketitle

\textbf{Quantum emitters hosted in crystalline lattices are highly attractive candidates  for quantum information processing \cite{Waldherr:2014kt}, secure networks  \cite{Togan:2010hi,DeGreve:2012iv}  and  nanosensing \cite{Mamin:2013eu, Staudacher:2013kn}. For many of these applications it is necessary to have control over single emitters with long spin coherence times. Such single quantum systems have been 
realized  using quantum dots \cite{Atature:2007dk}, colour centres in diamond \cite{Balasubramanian:2009fu}, dopants in nanostructures \cite{Siyushev:2014ie} and molecules \cite{Nothaft:2012eo}. More recently, ensemble emitters with spin dephasing times on the order of microseconds \cite{Koehl:2011fv, Soltamov:2012ey} and room-temperature optically detectable magnetic resonance \cite{Kraus:2013di} have been identified in  silicon carbide (SiC), a compound being highly compatible to up-to-date semiconductor device technology. So far however, the engineering of such spin centres in SiC on single-emitter level has remained elusive \cite{Baranov:2011ib}.  
Here, we demonstrate the control of spin centre density in ultrapure SiC over 8 orders of magnitude, from below $\mathbf{10^{9}}$ to above $\mathbf{10^{16} \, cm^{-3}}$ using neutron irradiation.  For a low irradiation dose, a fully photostable, room-temperature, near infrared (NIR) single photon emitter can clearly be isolated, demonstrating no bleaching even after $\mathbf{10^{14}}$ excitation cycles. Based on their spectroscopic fingerprints, these centres are identified as silicon vacancies, which 
can potentially be used as qubits \cite{Riedel:2012jq}, spin sensors \cite{Kraus:2013vf} and maser amplifiers \cite{Kraus:2013di}.}  

Silicon vacancy ($\mathrm{V_{Si}}$)-related defects in SiC can significantly exceed the performance of on-chip photonic networks and long-distance quantum communication systems, compared to many other  solid-state single photon emitters. In particular, the zero-phonon lines (ZPLs) of $\mathrm{V_{Si}}$-related defects in 4H, 6H and 3C polytypes of SiC present spectrally narrow features at NIR wavelengths $\lambda_{\mathrm{ZPL}} = 850 - 1200 \, \mathrm{nm}$. Rayleigh scattering losses in photonic structures are inversely proportional to the fourth power of the wavelength, giving almost one order of magnitude lower losses for these defects compared to the nitrogen-vacancy defect in diamond ($\lambda_{\mathrm{ZPL}} = 630  \, \mathrm{nm}$) \cite{Kurtsiefer:2000tk} or the carbon antisite-vacancy pair in SiC ($\lambda_{\mathrm{ZPL}} = 660 \, \mathrm{nm}$) \cite{Castelletto:2013el}. Similarly, scattering losses at interfaces and signal attenuation in optical fibers decrease with wavelength as well. Furthermore, $\mathrm{V_{Si}}$-related defects in SiC can be integrated with existing optoelectronic devices \cite{Fuchs:2013dz} and, in contrast to GaAs-based quantum dots \cite{Yuan:2001dp}, operate even at room temperature. 

\begin{figure*}[t]
\includegraphics[width=.93\textwidth]{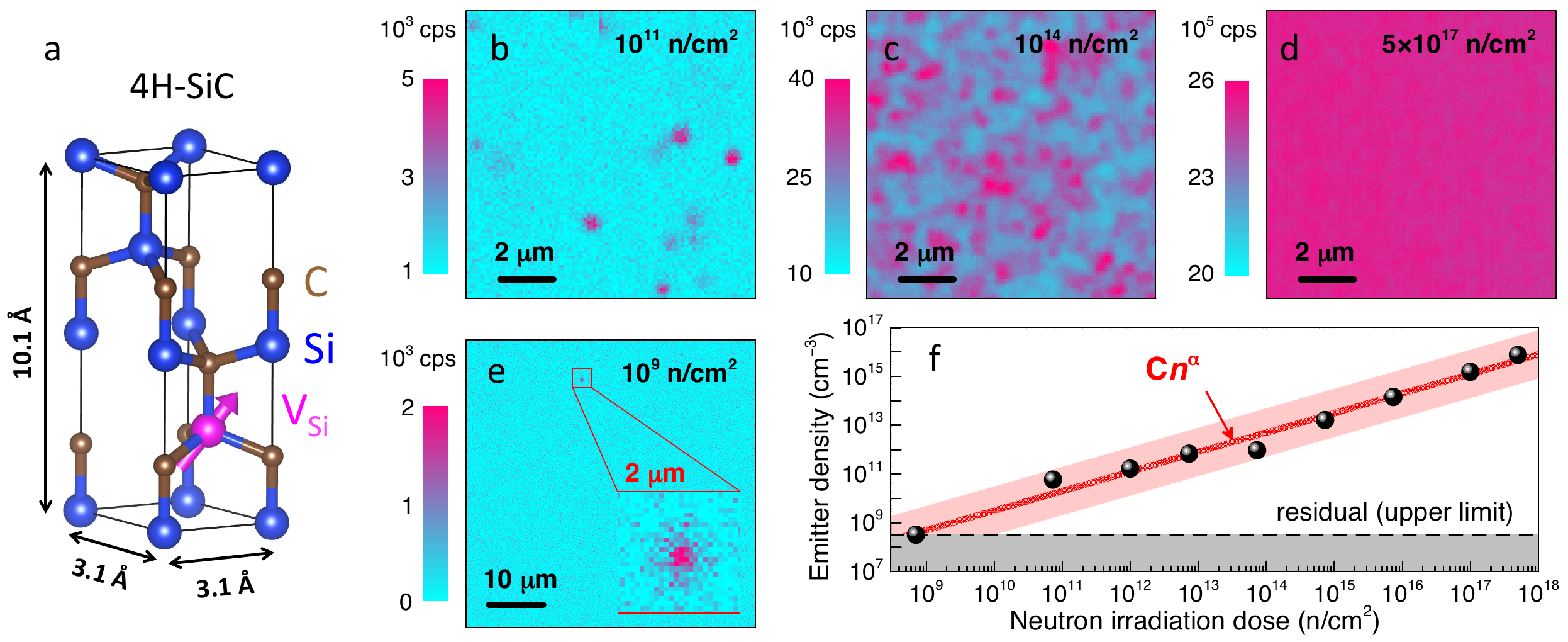}
\caption{Generation of $\mathrm{V_{Si}}$ defects in ultrapure 4H-SiC samples by neutron irradiation. (a) A scheme of the 4H-SiC unit cell with a single $\mathrm{V_{Si}}$ defect. (b)-(d) Confocal microscopy raster scans ($10 \times 10 \, \mathrm{\mu m^2}$) for different neutron irradiation doses: (b) $n =1 \times 10 ^{11} \, \mathrm{n/cm^2}$, (c) $n = 1 \times 10 ^{14} \, \mathrm{n/cm^2}$ and (d) $n = 5 \times 10 ^{17} \, \mathrm{n/cm^2}$. (e) A confocal microscopy raster scan ($50 \times 50 \, \mathrm{\mu m^2}$) with a single $\mathrm{V_{Si}}$ defect (shown in the inset) for $n = 1 \times 10 ^{9} \, \mathrm{n/cm^2}$. (f) Concentration of single photon emitters $\mathcal{N}$ as a function of the irradiation dose. The solid line is a fit to $\mathcal{N} = \mathcal{C} n^{\alpha}$ with $\alpha = 0.8$. } \label{fig1}
\end{figure*}

The 4H-SiC unit cell with single $\mathrm{V_{Si}}$ defect is shown in Fig.~\ref{fig1}(a). The dangling bonds of four C atoms with the absent Si atom result in formation of energy levels within the forbidden gap ($3.23 \, \mathrm{eV}$) of 4H-SiC \cite{Bockstedte:2003fn,Gali:2012jy}. In case of negatively-charged $\mathrm{V_{Si}}$, five electrons form a spin quadruplet ($S=3/2$) in the ground state \cite{Mizuochi:2002kl,Kraus:2013di}.   To excite these defects we use sub-band gap excitation of SiC at a laser wavelength of  $785 \, \mathrm{nm}$ ($h \nu = 1.58 \, \mathrm{eV}$), which is close to their optimal excitation wavelength \cite{Hain:2014tl}. At room temperature, the $\mathrm{V_{Si}}$ defects emit in the NIR spectral range from $800$ to $1100 \, \mathrm{nm}$. At cryogenic temperatures, two  distinct ZPLs at $\lambda_{\mathrm{ZPL}} = 862 \, \mathrm{nm}$ (V1) and $\lambda_{\mathrm{ZPL}} = 917 \, \mathrm{nm}$ (V2), associated with two different crystallographic sites in 4H-SiC, are  observed in the photoluminescence (PL) spectrum \cite{Sorman:2000ij}, which can be used as spectroscopic fingerprints of $\mathrm{V_{Si}}$.  

To control the $\mathrm{V_{Si}}$ density in a high-quality 110-$\mathrm{\mu m}$-thick 4H-SiC epitaxial layer \cite{Hain:2014tl}, we used neutron irradiation ($0.18 \, \mathrm{MeV}  < E_n  <  2.5 \, \mathrm{MeV}$) in a fission reactor. The irradiation dose was varied over  more than 8 orders of magnitude, from $10^{9}$ to $5 \times 10^{17} \, \mathrm{n / cm^{2}}$ (Fig.~\ref{fig1}). Some part of the generated $\mathrm{V_{Si}}$ defects are negatively charged due to the presence of residual N donors ($5.0 \times 10^{14} \, \mathrm{cm ^{-3}}$). Additionally, the neutron transmutation doping $\mathrm{^{30}Si}(\mathrm{n}, \gamma)\mathrm{^{31}Si} \rightarrow \mathrm{^{31}P + \beta^{-}}$ may play a role. 

\begin{figure}[ht]
\includegraphics[width=.47\textwidth]{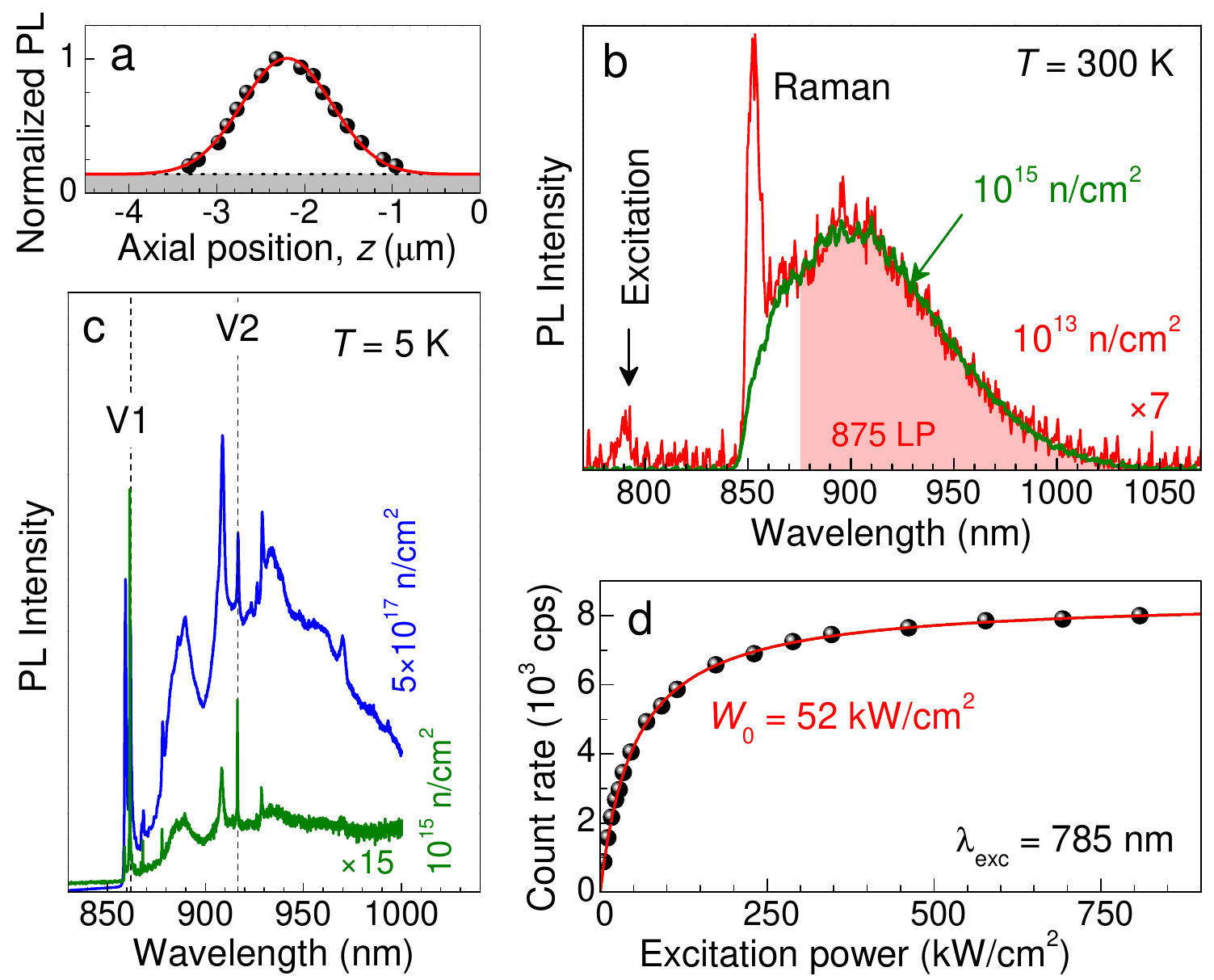}
\caption{NIR emission of a single $\mathrm{V_{Si}}$ defect in 4H-SiC. (a) PL line scan through a single $\mathrm{V_{Si}}$ defect. The top of the undoped 4H-SiC layer is placed at $z = 0 \, \mathrm{\mu m}$. Integrated PL is obtained with a LP filter at $875 \, \mathrm{nm}$. The solid line is a Gauss fit with a FWHM of $1.2 \, \mathrm{\mu m}$.  (b)  Room-temperature PL spectrum of a single $\mathrm{V_{Si}}$ defect ($n = 10 ^{13} \, \mathrm{n/cm^2}$) in comparison to a $\mathrm{V_{Si}}$ ensemble ($n = 10 ^{15} \, \mathrm{n/cm^2}$). A $850 \, \mathrm{nm}$ LP filter is used to suppress the excitation at $785 \, \mathrm{nm}$.  An additional LP filter $875 \, \mathrm{nm}$ can be used to suppress a Raman line (LO phonon). (c) Low-temperature PL spectra of $\mathrm{V_{Si}}$ defects for different irradiation doses. Two ZPLs, labeled as V1 and V2 (vertical dashed lines), are characteristic for two types of the $\mathrm{V_{Si}}$ defect in 4H-SiC.  (d) Photon count rate of a single $\mathrm{V_{Si}}$ defect as a function of excitation power density. The solid line is a fit to Eq.~(\ref{Power}) with $W_0 = 52 \, \mathrm{kW / cm^{2}}$. } \label{fig2}
\end{figure}

A PL confocal raster scan ($10 \times 10 \, \mathrm{\mu m^2}$) on a sample irradiated with a low dose of $n = 1 \times 10^{11} \, \mathrm{n / cm^{2}}$ is presented in Fig.~\ref{fig1}(b). The PL is detected in the spectral range from $875 \, \mathrm{nm}$ [owing to a longpass (LP) filter] to $1050 \, \mathrm{nm}$ [limited by the sensitivity of Si avalanche photodiodes (APDs)]. Four, nearly diffraction-limited spots [full width at half maximum (FWHM) of ca. $500 \, \mathrm{nm}$] are clearly seen in this scan. With rising irradiation dose to $n = 1 \times 10^{14} \, \mathrm{n / cm^{2}}$ the number of PL spots increases as well [Fig.~\ref{fig1}(c)]. For the highest irradiation dose of $n = 5 \times 10^{17} \, \mathrm{n / cm^{2}}$ single PL spots cannot be resolved any more and the PL spatial distribution becomes highly homogeneous [Fig.~\ref{fig1}(d)]. Remarkably, in the negligibly weak irradiated sample (dose of $1 \times 10^{9} \, \mathrm{n / cm^{2}}$) only one PL spot is found in the $50 \times 50 \, \mathrm{\mu m^2}$ raster scan [Fig.~\ref{fig1}(e)]. The single spots are also restricted along the optical axis as shown in Fig.~\ref{fig2}(a). Below we unambiguously prove that these intensity spots are due to the emission from single $\mathrm{V_{Si}}$ defects. 

In order to find the density of single $\mathrm{V_{Si}}$ photon emitters the following procedure is used. Up to the irradiation dose of $n = 1 \times 10^{14} \, \mathrm{n / cm^{2}}$ we directly  count the number of PL spots in the detection volume, given by the scanned area and the focus depth, the latter is about $1.2 \, \mathrm{\mu m}$ according to Fig.~\ref{fig2}(a). For this irradiation dose we also measure the integrated PL intensity collected from an area of about  $100 \, \mathrm{\mu m^2}$. This PL is used as a reference to calculate the emitter density in the strongly irradiated samples by comparing PL intensities. The results are presented in Fig.~\ref{fig1}(f). For the lowest irradiation dose the defect density is $\mathcal{N} = 3 \times 10^{8} \, \mathrm{cm^{-3}}$, which can be taken as the upper limit of residual $\mathrm{V_{Si}}$ concentration in our 4H-SiC sample. The concentration after the highest irradiation dose constitutes $\mathcal{N} = 7 \times 10^{15} \, \mathrm{cm^{-3}}$. The irradiation dose dependence follows quite well a polynomial scaling  $\mathcal{N}  \propto n^{0.8}$, as shown by the solid line in Fig.~\ref{fig1}(f).

To identify the type of generated defects, we measure the PL spectra for different irradiation doses. The spectrum from a single center is identical to the ensemble emission [Fig.~\ref{fig2}(b)]. 
Here we use a LP filter $850 \, \mathrm{nm}$ to suppress the excitation light at $785 \, \mathrm{nm}$ in the detection path. 
The LO phonon Raman line from 4H-SiC at $850 \, \mathrm{nm}$ is independent of the irradiation dose and hence is masked by the stronger PL band for  $n = 10 ^{15} \, \mathrm{n/cm^2}$. We therefore use an additional LP filter $875 \, \mathrm{nm}$  when investigating single centers.  The PL spectra recorded at low temperature ($T = 5 \, \mathrm{K}$) are presented in Fig.~\ref{fig2}(c). Two characteristic lines at $861.4 \, \mathrm{nm}$ and $916.3 \, \mathrm{nm}$ are clearly visible for different irradiation doses. These lines  coincide with the V1 and V2 ZPLs \cite{Sorman:2000ij}, proving that the PL originates from the $\mathrm{V_{Si}}$ defects in 4H-SiC. 

\begin{figure}[ht]
\includegraphics[width=.47\textwidth]{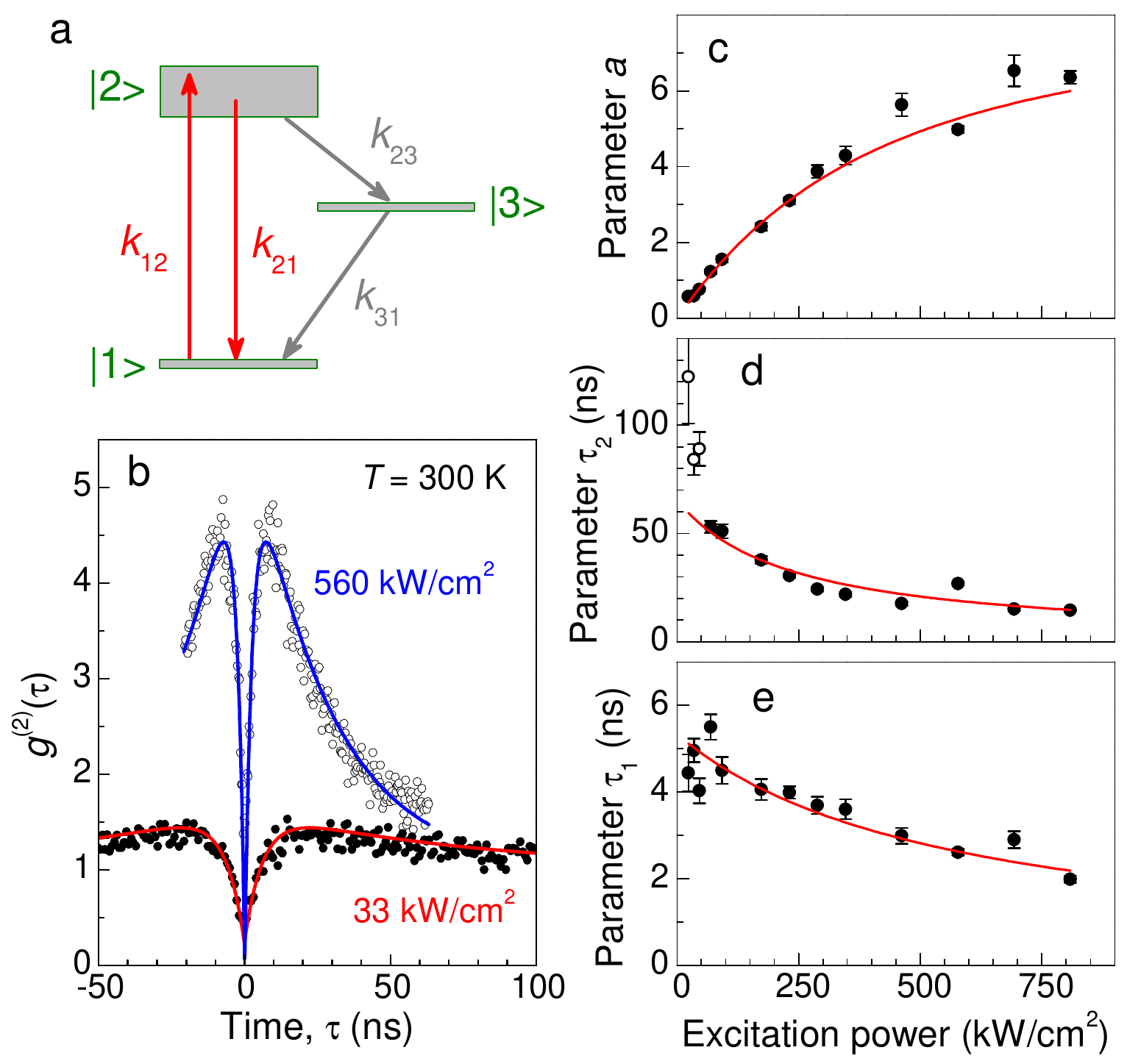}
\caption{Intensity correlation measurements at room temperature. (a) A three-level model of the $\mathrm{V_{Si}}$ defect with radiative recombination from the excited state $| 2 \rangle$ to the ground state $| 1 \rangle$ and nonradiative relaxation through the metastable state $| 3 \rangle$. (b) Correlation function $g^{(2)} (\tau)$ recorded at two different excitation power densities $W < W_0$ and $W \gg W_0$. The solid lines are fits to Eq.~(\ref{AntiBunch}). (c)-(e) Fit parameters $a$, $\tau_1$ and $\tau_2$ of the antibunching curve as a function of excitation power density. The solid lines are fits to the model presented in the text.} \label{fig3}
\end{figure}

As expected for single defect centers, the PL intensity $I$ saturates with increasing excitation power density $W$ [Fig.~\ref{fig2}(d)]. After subtracting the linear background contribution and APD dark counts, it follows 
\begin{equation}
 I (W) =    \frac{I_{max}}{1 + W_0 / W}  \,.
 \label{Power}
\end{equation}
Here, $W_0 = 52 \, \mathrm{kW / cm^{2}}$ is the saturation power density exposed to the sample, corresponding to the laser power $P_0 = 0.3 \, \mathrm{mW}$ at the objective entrance aperture. The saturation PL intensity varies slightly from spot to spot and in Fig.~\ref{fig3}(d) yields $I_{max} = 8.5 \times 10^3$ counts per second (cps).  

As a next step, we perform the Hanbury-Brown and Twiss (HBT) interferometry experiment, i.e., the time correlation measurement of photon detection by two APDs. This is a frequently used method to verify single photon emission \cite{Kurtsiefer:2000tk,Castelletto:2013el}. The second-order correlation functions $g^{(2)} (\tau)$, recorded over several hours for different $W$, are shown in Fig.~\ref{fig3}(b). The most important feature is the dip at zero time delay ($\tau = 0$). For the lowest excitation, we obtain $g^{(2)} (0) = 0.23 \pm 0.07 <  0.5$, which denotes clearly the non-classical behavior of a single photon emitter. Additionally to the anti-bunching for $| \tau | < 15 \, \mathrm{ns}$ there is also bunching for $| \tau | > 15 \, \mathrm{ns}$. In order to explain such a  behaviour at least three levels should be involved [Fig.~\ref{fig3}(a)].  

The second-order correlation function can be well described using 
\begin{equation}
 g^{(2)} (\tau) =   1 - (1 + a) e^{- | \tau | / \tau_1} + a  e^{- | \tau | / \tau_2} \,,
 \label{AntiBunch}
\end{equation}
as shown by the solid lines in Fig.~\ref{fig3}(b).  
The power dependencies of parameters $a$, $\tau_2$ and $\tau_1$ are presented in Figs.~\ref{fig3}(c)-(e), respectively. We use the same three-level model as for the colour centres in diamond \cite{Kurtsiefer:2000tk,Aharonovich:2010bj,Neu:2011ik}  to fit these dependencies. This model describes reasonably well the bunching amplitude $a (W)$ [the solid line in Fig.~\ref{fig3}(c)] and the anti-bunching decay time $\tau_1 (W)$  [the solid line in Fig.~\ref{fig3}(e)]. However, the relatively long bunching decay time $\tau_2 (W)$ for $W < 100 \, \mathrm{kW / cm^{2}}$  is not well reproduced within this model [the solid line in Fig.~\ref{fig3}(d)]. The fit for $\tau_2$ also provides significantly different transition rates $k_{ij}$ as that for $a$ and $\tau_1$.  A possible explanation is that a deshelving process of the metastable state $| 3 \rangle$ may occur under optical excitation \cite{Neu:2011ik}. 

In order to find the transition rates $k_{ij}$ of the three-level model in Fig.~\ref{fig3}(a), we take the limiting values for $W \rightarrow 0$ and $W = 800 \,  \mathrm{kW / cm^{2}} \gg W_0$ as an approximation for $W \rightarrow \infty$. We obtain $\tau_1 (0) = 1/(k_{21} + k_{23}) = 5.3 \, \mathrm{ns}$, $\tau_2 (\infty) = 1/(k_{23} + k_{31}) = 14.5 \, \mathrm{ns}$, $a (\infty) = k_{23} / k_{31} = 6.4$, and the corresponding lifetimes are summarised in table~\ref{Rates}.   Remarkably, the value for $\tau_1$ agrees well with the PL decay time of $6.1 \, \mathrm{ns}$ observed in time-resolved experiments \cite{Hain:2014tl} and remarkably longer than that of the band-to-band transition in semiconductor nanostructures \cite{Yakovlev:2000ez}. The excitation rate of $\mathrm{V_{Si}}$ is proportional to the laser power density $k_{12} = \sigma W$, where the absorption cross section $\sigma$ can be calculated from the saturation behaviour of  Fig.~\ref{fig2}(d) as $\sigma  = (h \nu / T W_0) (k_{23} k_{31} + k_{21} k_{31}) / (k_{23} + k_{31})$. Here, we take for the transmission coefficient at the SiC surface $T = 0.81$ and the calculated value for $\sigma$  is also presented in table~\ref{Rates}.  

\begin{table}[htdp]
\caption{Transition rates $k_{ij}$ and absorption cross section $\sigma$ governing the population dynamics of $\mathrm{V_{Si}}$. }
\begin{center}
\begin{tabular}{|c|c|c|c|}
$\,\,\,\,\, 1 / k_{21} \,\,\,\,\,$ & $\,\,\,\,\, 1 / k_{23} \,\,\,\,\,$ & $\,\,\,\,\, 1 / k_{31} \,\,\,\,\,$ & $\,\,\,\,\,\,\,\, \sigma \,\,\,\,\,\,\,\,$ \\ \hline
$7.6 \, \mathrm{ns}$ & $16.8 \, \mathrm{ns}$  & $107 \, \mathrm{ns}$  &  $1.5 \times 10^{-16} \, \mathrm{cm^2}$  \\
\end{tabular}
\end{center}
\label{Rates}
\end{table}

\begin{figure}[t]
\includegraphics[width=.47\textwidth]{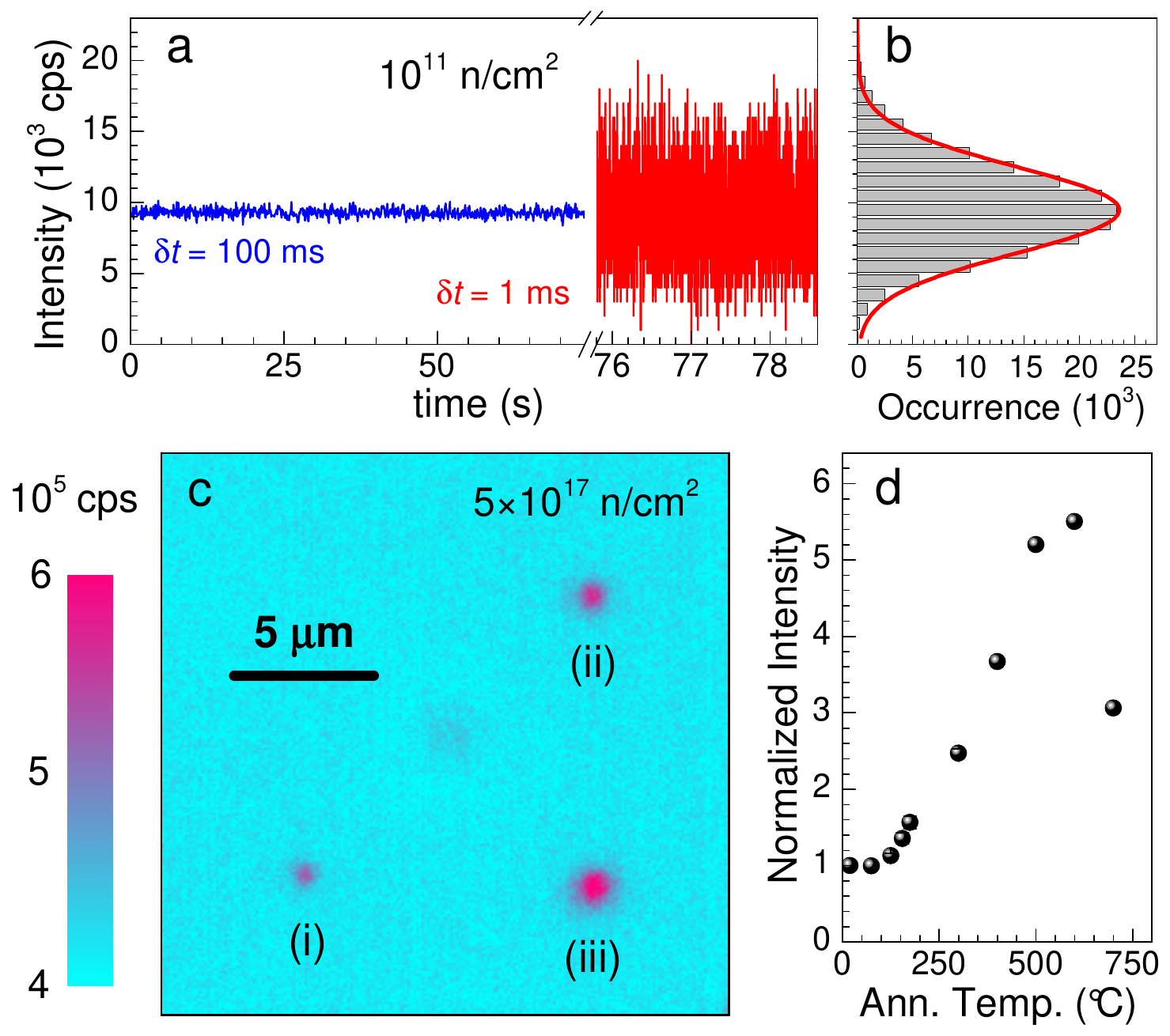}
\caption{Photostability of $\mathrm{V_{Si}}$ defects. (a) PL time traces with sampling bins $\delta t =  100 \, \mathrm{ms}$ and  $\delta t =  1 \, \mathrm{ms}$, obtained on a single  $\mathrm{V_{Si}}$ defect under excitation power density of  $1400 \, \mathrm{kW / cm^{2}}$. (b)  Corresponding count rate histogram for $\delta t =  1 \, \mathrm{ms}$. The solid line is a Gauss fit. (c) A confocal microscopy raster scan ($20 \times 20 \, \mathrm{\mu m^2}$) recorded after laser illumination of different spots during 120 minutes: spot (i) $P = 0.5 \, \mathrm{mW}$, spot (ii) $P = 1.5 \, \mathrm{mW}$ and spot (iii) $P = 3.5 \, \mathrm{mW}$.  (d) Relative change of the $\mathrm{V_{Si}}$ PL intensity as a function of annealing temperature. The $\mathrm{V_{Si}}$ density before annealing is $\mathcal{N} = 7 \times 10^{15} \, \mathrm{cm^{-3}}$. } \label{fig4}
\end{figure}

Photostability is an important characteristic of a single photon emitter. The PL time traces of a single $\mathrm{V_{Si}}$ defect are shown in  Fig.~\ref{fig4}(a).  For a sampling bin to $\delta t =  100 \, \mathrm{ms}$ the count rates remain constant over minutes. In order to examine the photostability on a shorter time scale, the sampling bin is reduced to  $\delta t =  1 \, \mathrm{ms}$. The number of detected photons per sampling bin is 10 in this case, and the time trace demonstrates statistical fluctuations without any indication of blinking [Fig.~\ref{fig4}(b)]. We have investigated a single $\mathrm{V_{Si}}$ emitter over more than one week under continuous excitation and did not observe photobleaching. Assuming that the excitation occurs on average every $10 \, \mathrm{ns}$, this corresponds to $10^{14}$ excitation cycles. 

Finally we found that in the highly irradiated sample the PL intensity increases locally upon laser illumination. To demonstrate this effect, the laser of different intensities was focussed sequently on three different spots and remained there for 120 minutes, respectively. A  confocal raster scan, performed at low laser power after such a procedure, demonstrates clearly a PL enhancement for each spot, as shown in Fig.~\ref{fig4}(c). For the highest laser power [spot (iii) in Fig.~\ref{fig4}(c)] this enhancement is ca. 25\%. Remarkably, the generated pattern of Fig.~\ref{fig4}(c) preserves at least one day.  

Our interpretation is that the focused laser beam locally heats the sample, resulting in atomic displacements and thus in disappearance or/and transformation of some other types of intrinsic defects, generated upon neutron irradiation. These defects may serve as non-radiative recombination channels or as charge traps, switching off the $\mathrm{V_{Si}}$ defects in close proximity.  In order to corroborate this explanation, we perform complementary annealing experiments by increasing stepwise the temperature and monitoring the PL intensity after each step. The results are summarized in Fig.~\ref{fig4}(d), showing the overall PL enhancement by a factor of $5.5$. This corresponds to $\mathcal{N} = 3.9 \times 10^{16} \, \mathrm{cm^{-3}}$ and demonstrates that $\mathrm{V_{Si}}$ defects can be created at high density in a controlled manner, as required, for instance, for the implementation of a SiC maser \cite{Kraus:2013di}. 

In our experiments, we precisely control the concentration of $\mathrm{V_{Si}}$ defects in ultrapure 4H-SiC down to single defect level. This approach can be used to deterministically incorporate these atomic-scale defects in electronic \cite{Fuchs:2013dz} and photonic structures \cite{Calusine:2014gv} as well as in nanocrystals. Together with their extremely narrow optical resonances (on the order of  $10 \, \mathrm{pm}$ at low temperature \cite{Riedel:2012jq}) and recently demonstrated optically-detected spin resonances at ambient conditions \cite{Kraus:2013vf}, our results open exciting opportunities for various quantum applications with spin-photon interface.  


\section*{Methods}

\subsection*{Samples}

The 4H-SiC sample has been purchased from CREE. A high purity (residual nitrogen doping below $5.0 \times 10^{14} \, \mathrm{cm ^{-3}}$) layer of $110 \, \mathrm{\mu m}$ thickness was epitaxially grown on a 2-inch n-type 4H-SiC wafer. 
The layer is covered by a 5-$ \mathrm{\mu m}$-thick n-type 4H-SiC layer and a 2-$ \mathrm{\mu m}$-thick p-type 4H-SiC layer. The wafer was diced in  $4 \, \mathrm{mm} \times 2 \, \mathrm{mm}$ pieces, which were then irradiated in a TRIGA Mark-II nuclear reactor, with neutron energies in the range of $0.18\,$MeV$ < E_n < 2.5\,$MeV. 

The sample with the highest neutron irradiation dose of $5 \times 10^{17} \, \mathrm{n / cm^{2}}$  was thermally annealed in several steps from $125^{\circ}\mathrm{C}$ to $700^{\circ}\mathrm{C}$ for a time of 90 minutes, respectively. The heating was either performed on a heat stage ($125^{\circ}\mathrm{C}$--$200^{\circ}\mathrm{C}$) or in an oven ($300^{\circ}\mathrm{C}$--$700^{\circ}\mathrm{C}$).

\subsection*{Experimental setup}

Samples were investigated with a home-build confocal microscope and a cw 785~nm laser was used for excitation. A three dimensional piezo unit (nPoint) was used to move the SiC sample in lateral and axial directions. The excitation beam was focused onto the samples by a high aperture ($\mathrm{NA} = 1.49$) oil immersion microscope objective (UAPON 100XOTIRF, Olympus). Collimated optical response of the sample was collected by the same objective and guided through a 30 or 75~$\mathrm{\mu m}$ pinhole (Thorlabs) followed by a 850 and 875~nm LP filter (Edmund Optics). Time correlated single photon counting (TCSPC) was recorded by a HBT setup consisting of two APDs (Count-100C-FC, Laser-components GmbH) with a quantum efficiency of about 0.3 at the signal wavelength and less than 100~cps in the dark and a 16-channel photon correlator card (DPC-230, Becker\&Hickl GmbH) with a time resolution of at least 165~ps.

The low temperature PL spectra were measured at 5~K in a cryostat (MicrostatHe, Oxford Instruments) built into a confocal Raman spectrometer (LabRAM HR, Horiba). The excitation wavelength was 633~nm.


\section*{Acknowledgments}
This work has been supported by the German Research Foundation (DFG) under grant AS 310/4 and by the Bavarian State Ministry of Science, Research, and the Arts within the Collaborative Research Network ÒSolar Technologies Go Hybrid. We thank R.~Bergmann and M.~Villa for assistance with neutron irradiation and N.~Wolf for help with annealing. 



\end{document}